\documentstyle[epsf,twocolumn,seceq]{orb2001}

\newcommand{\ee}{{\rm e}}
\newcommand{\ii}{{\rm i}}

\title
{
Magnetism and Superconductivity in a Two-band Hubbard Model in Infinite Dimensions
}

\def\runtitle
{
Magnetism and Superconductivity in a Two-band Hubbard Model in Infinite Dimensions
}

\author
{
Yoshiaki {\sc \=Ono}* and Kazuhiro {\sc Sano}$^{1}$**
}

\def\runauthor
{
Yoshiaki {\sc \=Ono} and Kazuhiro {\sc Sano}
}

\inst
{
Department of Physics, Nagoya University, Nagoya 464-8602, Japan\\ 
$^{1}$Department of Physics Engineering, Mie University, Tsu, Mie 514-8507, Japan
}

\abst
{
We study a two-band Hubbard model using the dynamical mean-field theory combined with the exact diagonalization method. At the electron density $n=2$, a transition from a band-insulator to a correlated semimetal occurs when the on-site Coulomb interaction $U$ is varied for a fixed value of the charge-transfer energy $\Delta$. At low temperature, the correlated semimetal shows ferromagnetism or superconductivity. With increasing doping $|n-2|$, the ferromagnetic transition temperature rapidly decreases and finally becomes zero at a critical value of $n$. The second-order phase transition occurs at high temperature, while a phase separation of ferromagnetic and paramagnetic states takes place at low temperature. The superconducting transition temperature gradually decreases and finally becomes zero near $n=1$ ($n=3$) where the system is Mott insulator which shows antiferromagnetism at low temperature. 
}

\kword
{
two-band Hubbard model, metal-insulator transition , ferromagnetism, antiferromagnetism, superconductivity, dynamical mean-field theory, infinite dimensions, exact diagonalization method
}

\begin{document}
\sloppy
\maketitle

\footnotetext{* E-mail: c42545a@nucc.cc.nagoya-u.ac.jp}
\footnotetext{** E-mail: sano@phen.mie-u.ac.jp}

\section{Introduction}

The two-band Hubbard model has been extensively studied as a simple model simulating transition-metal compounds and high-$T_c$ superconductors, which is characterized by two parameters: the on-site Coulomb interaction $U$ at the $d$-orbital and the charge-transfer energy $\Delta$ between the $d$- and $p$-orbitals. Recently, several authors have studied the model using the dynamical mean-field theory (DMFT) which becomes exact in infinite spatial dimensions \cite{Georges, Caffarel, Mutou, Watanabe, Ono1, Ono2, Ohashi}. The Mott metal-insulator transition was found to occur at $n=1$ (or $n=3$) \cite{Georges, Mutou, Ono1, Ono2, Ohashi}, where $n$ is the total electron number per unit cell and given by the sum of $p$- and $d$-electron numbers: $n=n_p+n_d$. Antiferromagnetism \cite{Watanabe} and superconductivity \cite{Georges, Caffarel, Ono1} were also observed in this model. However, the ferromagnetism was not discussed there.

Recent findings of the high-temperature ferromagnetism in doped CaB$_6$ and the high-temperature superconductivity in MgB$_2$ have simulated a renewed interest in strong electron correlation in the multi-band systems. 
Therefore, we study the ferromagnetism and the superconductivity in the two-band Hubbard model using the DMFT combined with the exact diagonalization (ED) method \cite{Caffarel, Ono1}. We especially focus on the electronic states near $n=2$ where the system is semimetal or band-insulator in the non-interacting case.

\section{Formulation}

The  two-band  Hubbard  model is given by 
\begin{eqnarray} 
H &=& \sum_{i,j,\sigma} (t_{ij} d_{i\sigma}^{\dagger} p_{j\sigma}+h.c.)
      + \epsilon_p\sum_{j,\sigma}p_{j\sigma}^{\dagger} p_{j\sigma}
      \nonumber \\
  &+& \epsilon_d\sum_{i,\sigma}d_{i\sigma}^{\dagger} d_{i\sigma}
     + U\sum_{i}d^{\dagger}_{i\uparrow}d_{i\uparrow} 
                d^{\dagger}_{i\downarrow}d_{i\downarrow}, 
  \label{MODEL}
\end{eqnarray} 
where $d^{\dagger}_{i\sigma}$ and $p_{j\sigma}^{\dagger}$ are creation operators for an electron with spin $\sigma$ in the $d$- orbital at site $i$ and in the $p$- orbital at site $j$, respectively. 
The charge-transfer energy $\Delta$ is defined by 
$\Delta=\epsilon_{p}-\epsilon_{d}$. 
In eq.(\ref{MODEL}), we assume that $p$- and $d$-orbitals are on different sub-lattices of a bipartite lattice. 

In the limit of infinite dimensions, the self-energy becomes purely site-diagonal and the DMFT becomes exact. 
The local Green's function for the $d$-electron, $D_{\sigma}(\tau-\tau')=-\langle Td_{\sigma}(\tau)d^{\dagger}_{\sigma}(\tau')\rangle$, 
can be given by the impurity Green's function of an effective single impurity Anderson model, 
\begin{eqnarray} 
H_{\rm And}
  &=& \sum_{\sigma} \varepsilon_{f\sigma} f^{\dagger}_\sigma f_\sigma 
         + U f^{\dagger}_{\uparrow}f_{\uparrow} 
             f^{\dagger}_{\downarrow}f_{\downarrow} \nonumber \\ 
  &+& \sum_{k,\sigma} \varepsilon_{k\sigma} c^{\dagger}_{k\sigma} c_{k\sigma} 
         + \sum_{k,\sigma}V_{k\sigma}(f^{\dagger}_{\sigma}c_{k\sigma}
          +c^{\dagger}_{k\sigma}f_{\sigma}), 
  \label{AND}
\end{eqnarray} 
where $\varepsilon_{f\sigma}$ is the impurity level and $\varepsilon_{k\sigma}$ are energies of conduction electrons hybridized with the impurity by $V_{k\sigma}$. 
In the model eq. (\ref{AND}), the non-interacting impurity Green's function, 
$
{\cal G}_{0\sigma}(\ii\omega_n) = \left(\ii\omega_n-\varepsilon_{f\sigma} 
 - \sum_k \frac{V_{k\sigma}^2}{\ii\omega_n-\varepsilon_{k\sigma}}\right)^{-1}, 
$
includes effects of the interaction at all the sites except the impurity site and is determined self-consistently so as to satisfy the self-consistency equation. 

For simplicity, the calculations in this paper are restricted to the Bethe lattice with the connectivity $z$ and the hopping $t_{ij}=\frac{t_{pd}}{\sqrt{z}}$, and we set $t_{pd}=1$. In the limit $z=\infty$, the self-consistency equations for the local  Green's functions are given by\cite{Georges}
\begin{equation} 
\begin{array}{rcl}
{\cal G}_{0\sigma}(\ii\omega_n )^{-1}
     &=& \ii\omega_n+\mu-\varepsilon_d-t_{pd}^2 P_{\sigma}(\ii\omega_n), 
\nonumber \\
P_{\sigma}(\ii\omega_n )^{-1}
     &=& \ii\omega_n+\mu-\varepsilon_p-t_{pd}^2 D_{\sigma}(i\omega_n), 
 \label{SCE}
\end{array}
\end{equation} 
where $P_{\sigma}(i\omega )$ is the local Green's function for the $p$-electron and $\mu $ is the chemical potential.

To solve the impurity Anderson model eq.(\ref{AND}) we use the exact diagonalization of a finite-size cluster with $\varepsilon_{k\sigma}$ and $V_{k\sigma}$ for $k=2,3,...,N_s$ (ED method) \cite{Caffarel}. 
First, we assume $4N_s-2$ parameters: $\{\varepsilon_{f\sigma},\varepsilon_{k\sigma},V_{k\sigma}\}_{\rm old}$. 
Next, we diagonalize the finite-size cluster with these parameters, and calculate the local (impurity) Green's function $D_{\sigma}(\ii\omega_n)$. 
Finally, we calculate ${\cal G}_{0\sigma}(\ii\omega_n)$ through the self-consistency equations (\ref{SCE}), and determine the new parameters 
$\{\varepsilon_{f\sigma},\varepsilon_{k\sigma},V_{k\sigma}\}_{\rm new}$ 
so as to make the non-interacting Green's function with finite $N_s$, ${\cal G}^{N_s}_{0\sigma}(\ii\omega_n)$, as close to ${\cal G}_{0\sigma}(\ii\omega_n)$ as possible, namely, so as to minimize $\chi^2$ defined by 
$
\chi^2= \sum_{\omega_n}|{\cal G}_{0\sigma}(\ii\omega_n)^{-1}
            - {\cal G}^{N_s}_{0\sigma}(\ii\omega_n)^{-1}|^2, 
$
where 
$
{\cal G}^{N_s}_{0\sigma}(\ii\omega_n)=(\ii\omega_n-\varepsilon_{f\sigma}
  - \sum_{k=2}^{N_s}\frac{V_{k\sigma}^2}
                         {\ii\omega_n-\varepsilon_{k\sigma}} )^{-1}.
$
This process is iterated until the solutions converge.

\section{Metal-Insulator Transition at $n=2$}

First, we consider the paramagnetic state at zero temperature for $n=n_d+n_p=2$. In the non-interacting case $U=0$, the system is a band-insulator for $\Delta\ne 0$ while it is a semimetal for $\Delta=0$. Within the restricted Hartree-Fock approximation (HFA), the energy gap is given by $\Delta-\frac{U n_d}{2}$. Then the system is metallic for $U= 2\Delta$, otherwise it is insulating. We note that $n_d=n_p=1$ for $U=2\Delta$ due to the particle-hole symmetry.

\begin{figure}[t]
\vspace{5mm}
\begin{center}
\leavevmode
\epsfxsize=6cm
    \epsffile{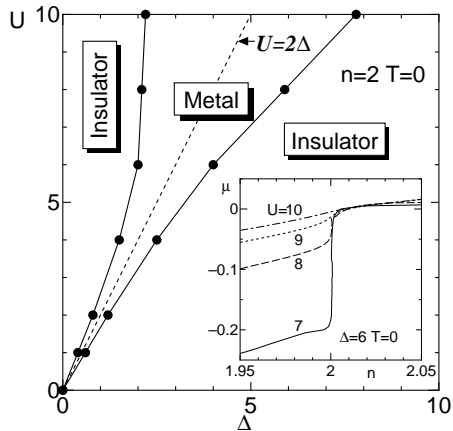}
\caption{
The phase boundary separating the metallic and insulating regimes as a function of $U$ and $\Delta$ at $n=2$ and $T=0$. The inset shows the chemical potential as functions of $n$ for $U=7, 8, 9, 10$ at $\Delta=6$ and $T=0$ calculated from the DMFT (ED method) with $N_s=8$.
}
\label{fig1}
\end{center}
\end{figure}
In the inset in Fig.~\ref{fig1}, we show the chemical potential $\mu$ as functions of $n$ for $U=7, 8, 9, 10$ at $\Delta=6$ and $T=0$ calculated from the ED method with the system size $N_s=8$. 
In this calculation, the solution is restricted to the paramagntic state with $\varepsilon_{f\sigma}=\varepsilon_{f}$, $\varepsilon_{k\sigma}=\varepsilon_{k}$ and $V_{k\sigma}=V_{k}$ for $k=2,3,...,N_s$. 
When $U$ increases from $U=0$, the discontinuity in the chemical potential at $n=2$ decreases and finally becomes zero at a critical value, where a transition from the band-insulator to the correlated semimetal occurs \cite{Georges}. The critical values for the metal-insulator transition are plotted in Fig.~\ref{fig1}. In contrast to the restricted HFA, the metallic state is found in the wide parameter region due to a correlation effect considered in the DMFT.

\section{Ferromagnetism}

\begin{figure}[t]
\vspace{5mm}
\begin{center}
\leavevmode
\epsfxsize=6cm
    \epsffile{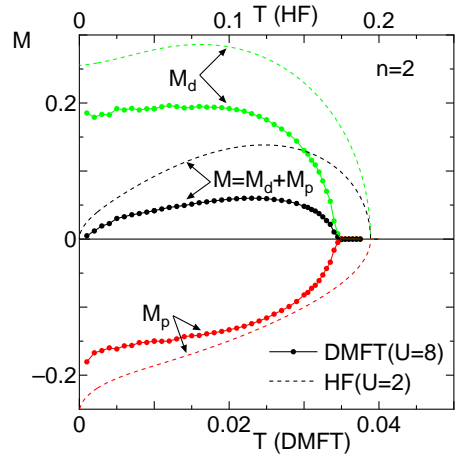}
\caption{
The magnetization for the $d$-electron $M_d$, that for the $p$-electron $M_p$ and the total magnetization $M=M_d+M_p$ as functions of the temperature $T$ at $n=2$, obtained from the ED method (closed circles) for $U=8$ and $\Delta=4$ with $N_s=6$ and from the Hartree-Fock approximation (dashed lines) for $U=2$ and $\Delta=1$. 
}
\label{fig2}
\end{center}
\end{figure}
At low temperature, the correlated semimetal mentioned above becomes unstable compared to a ferromagnetic state. In Fig.~\ref{fig2} we plotted the magnetization for the $d$-electron $M_d$, that for the $p$-electron $M_p$ and the total magnetization $M=M_d+M_p$ as functions of the temperature $T$ at $n=2$ for $U=8$ and $\Delta=4$ calculated from the ED method with $N_s=6$. As shown in Fig.~\ref{fig2}, $M_d$ and $M_p$ have opposite sign to each other. In the low temperature limit, both of $M_d$ and $M_p$ become constant, while the sum of them $M$ becomes zero. The feature of the ferromagnetism from the DMFT is similar to that from the HFA as shown in Fig.~\ref{fig2}. However, the transition temperature $T_c$ from the HFA is much higher than that from the DMFT (see also Fig.~\ref{fig4}).

\begin{figure}[t]
\vspace{5mm}
\begin{center}
\leavevmode
\epsfxsize=6cm
    \epsffile{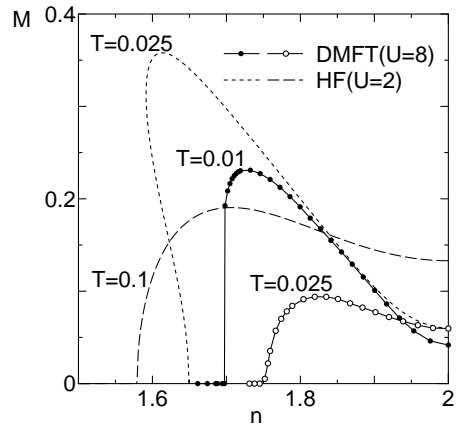}
\caption{
The total magnetization $M$ as a function of the electron number $n$, obtained from the DMFT for $U=8$ and $\Delta=4$ at $T=0.01$ (closed circles), $0.025$ (open circles), and from the HF approximation for $U=2$ and $\Delta=1$ at $T=0.025$ (dotted line), $0.1$ (dashed line). 
}
\label{fig3}
\end{center}
\end{figure}
Fig.~\ref{fig3} shows the total magnetization $M$ as a function of the electron number $n$. When $n$ decreases, $M$ continuously becomes zero at a critical value of $n$ for high temperatures (see for $T=0.025$), while it discontinuously becomes zero for low temperatures (see for $T=0.01$). The similar properties are also observed within the HFA as shown in Fig.~\ref{fig3}. At low temperatures, however, the HFA also predicts a metastable state where $M$ decreases with increasing $n$ and continuously becomes zero at a critical $n$ (see for $T=0.025$). By calculating the thermodynamic potential, we find that the phase separation of the ferromagnetic state and the paramagnetic state occurs at the low temperatures as shown in Fig.~\ref{fig4}. The possible metastable state and the phase separation within the ED method will be reported elsewhere.

\begin{figure}[t]
\vspace{5mm}
\begin{center}
\leavevmode
\epsfxsize=6.2cm
    \epsffile{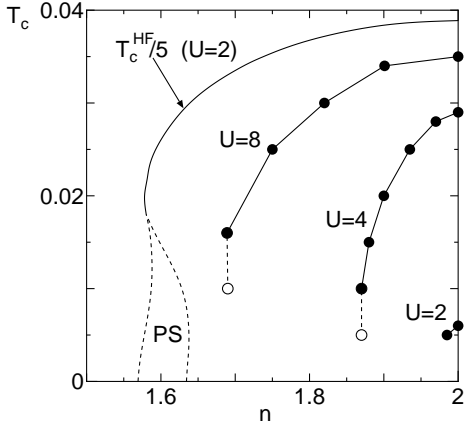}
\caption{
The transition temperature $T_c$ for the ferromagnetism as a function of the electron number $n$, obtained from the DMFT (closed and open circles) for $U=2\Delta=8, 4, 2,$ and from the HF approximation (solid and dotted lines) for $U=2\Delta=2$.
}
\label{fig4}
\end{center}
\end{figure}
Fig.~\ref{fig4} shows the transition temperature for the ferromagnetism $T_c$ as a function of $n$ for several values of $U(=2\Delta)$. $T_c$ monotonically decreases with decreasing $n$. The closed circles show the second-order phase transition, while the open circles show the discontinuous transition as seen in Fig.~\ref{fig3}. Within the HFA, the second-order phase transition occurs at the high temperature (solid line), while the phase separation occurs at the low temperature (area between the dotted lines).

\section{Superconductivity}

Finally, we discuss the superconductivity \cite{Georges,Caffarel,Ono1}. 
The on-site paring susceptibility $\chi_p$ is given by \cite{Georges}
\begin{eqnarray}
\chi_p &=& \frac{1}{N}\int^{\beta}_0 d\tau \sum_{ij}\langle T d_{i\uparrow}(\tau) d_{i\downarrow}(\tau)d_{j\downarrow}^{\dagger}(0) d_{j\uparrow}^{\dagger}(0)\rangle 
\nonumber \\
&=&T\sum_{\nu,\nu'}[\alpha^{-1/2} \{ I- \Lambda \}^{-1} \cdot \Lambda \cdot \alpha^{-1/2}  ]_{\nu,\nu'},
\end{eqnarray} 
where
$
[ \Lambda ]_{\nu,\nu'}=t_{pd}^4 |P(\ii\nu) | [\tilde{\chi}_{\rm loc}]_{\nu,\nu'} | P(\ii\nu') |, 
$
and 
$ 
[ \alpha ]_{\nu,\nu'}=t_{pd}^4 |P(\ii\nu)|^2 \delta_{\nu,\nu'}. 
$
Here, $\tilde{\chi}_{\rm loc}$ is the local paring susceptibility on a $d$-orbital given by
\begin{eqnarray}
        [\tilde{\chi}_{\rm loc}]_{\nu,\nu'}=T^2
\int^{\beta}_0 d\tau_1 \int^{\beta}_0 d\tau_2 \int^{\beta}_0 d\tau_3 \int^{\beta}_0 d\tau_4 \ee^{i\nu (\tau_1 - \tau_2 )}
\nonumber \\
 \times  \ee^{i\nu' (\tau_3 - \tau_4 )}
  \langle T d_{\uparrow}(\tau_1) d_{\downarrow}(\tau_2)d_{\downarrow}^{\dagger}(\tau_3) d_{\uparrow}^{\dagger}(\tau_4)\rangle. \ \
  \label{XLOC}
\end{eqnarray} 
To calculate $\tilde{\chi}_{\rm loc}$ within the exact diagonalization method, we use a spectral representation of r.h.s. in eq.(\ref{XLOC})  by inserting a complete set of eigenstates $|i\rangle$. 
When the largest eigenvalue of $\Lambda$ approaches unity, the paring susceptibility diverges. It signals the transition into the superconducting state from the normal state.

\begin{figure}[t]
\vspace{5mm}
\begin{center}
\leavevmode
\epsfxsize=6.5cm
    \epsffile{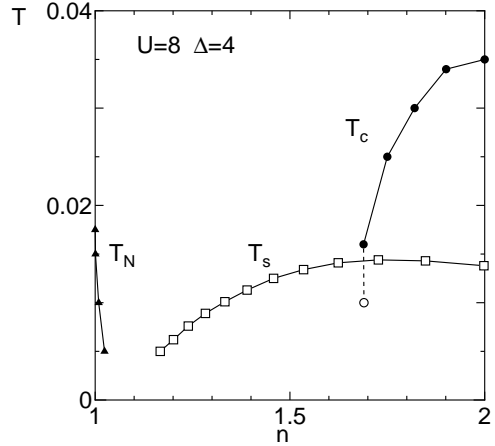}
\caption{
Transition temperatures for the ferromagnetism $T_c$, the singlet superconductivity $T_c$ and the antiferromagnetism $T_N$, obtained from the DMFT for $U=8$ and $\Delta=4$. 
}
\label{fig5}
\end{center}
\end{figure}
In Fig.~\ref{fig5}, the transition temperature for the singlet superconductivity $T_s$ is plotted as a function of $n$ for $U=8$ and $\Delta=4$, obtained from the ED method with $N_s=4$ \cite{Ono1}. We also plotted the transition temperature for the ferromagnetism $T_c$ calculated in $\S$4 together with that for the antiferromagnetism $T_N$ calculated from the ED method with $N_s=6$.

\section{Discussion}

We have obtained a phase diagram including the metal-insulator transition, the ferromagnetism, the antiferromagnetism and the superconductivity using the DMFT combined with the ED method, for the two-band Hubbard model with typical values of the parameters $U$ and $\Delta$. More systematic calculations for various values of the parameters are now under the way. 

Effects of the hopping integrals between the $p$-$p$ and the $d$-$d$ orbitals, which are not considered in the present study but are not negligible in actual compounds, make important contribution to the transition temperatures. Such effects and the detailed analysis of the mechanism for the superconductivity will be reported in a future publication.




\begin{thebibliography}{}


\bibitem{Georges}
A. Georges, G. Kotliar and W. Krauth: Z. Phys. {\bf B 92} (1993) 313.

\bibitem{Caffarel}
M. Caffarel and W. Krauth: Phys. Rev. Lett. {\bf 72} (1994) 1545. 

\bibitem{Mutou}
T. Mutou, H. Takahashi and D. S. Hirashima: J. Phys. Soc. Jpn. {\bf 66} (1997) 2781. 

\bibitem{Watanabe}
H. Watanabe and S. Doniach: Phys. Rev. B \textbf{57} (1998) 3829.

\bibitem{Ono1}
Y. \=Ono and K. Sano: J. Phys. Chem. Solids \textbf{62} (2001) 285.

\bibitem{Ono2}
Y. \=Ono, R. Bulla and A. C. Hewson, Eur. Phys. J. B \textbf{19} (2001) 375. 

\bibitem{Ohashi}
Y. Ohashi and Y. \=Ono: J. Phys. Soc. Jpn. {\bf 70}, No. 10 (2001), in press.







\end{thebibliography}
\end{document}